\documentclass[epj]{svjour}

\usepackage{graphicx}
\usepackage{fancyhdr}
\def\makeheadbox{{
 }}
\def\mytitle{My title} 
\def\myauthors{My name}  
\def\mytype{My type of session}
\def\mysession{My session}

\def\mytitle{Search of Squark/Gluino Production in CDF} 
\def\myauthors{Monica D'Onofrio}    
\def\mytype{Contributed Talk}    
\def\mysession{Colliders - SUSY Phenomenology}

\newcommand{\ETM}       {\rm E_T\hspace{-2.4ex}/\hspace{1.2ex}}
\newcommand{\MET}       {$\rm E_T\hspace{-2.4ex}/\hspace{1.2ex}~ $}
\newcommand{\gl}        {\tilde{g}}
\newcommand{\sq}        {\tilde{q}}

\newcommand{\mGeVcc}    {\mathrm{GeV/c}^2}

\begin{document}
\title{Inclusive Search for Squarks and Gluinos Production at CDF}
\subtitle{}
\author{Monica D'Onofrio\inst{}
\thanks{\emph{Email:} donofrio@fnal.gov}%
}                     
%
%
\institute{Institut de Fisica d'Altes Energies (IFAE), Barcelona}
%
\date{}
\abstract{
We present preliminary results on a search for squarks and gluinos in proton-antiproton
collisions with a center-of-mass energy of 1.96 TeV and based on 
1.1 fb$^{-1}$ of data collected by the CDF detector in the Tevatron Run II.
Events with multiple jets of hadrons and large missing transverse energy in the 
final state are studied within the framework of minimal supergravity and assuming 
R-parity conservation. No excess with respect to Standard Model predictions is 
observed and new limits on the gluino and squark masses are extracted. 
} 
\maketitle
%

\section{Introduction}
\label{intro}

Supersymmetry  (SUSY)~\cite{super} is $\,$ regarded $\,$ as $\,$ one of the most 
promising $\,$ fundamental $\,$ theories to $\,$ describe physics at 
arbitrarily high energies beyond the Standard Model (SM). 
Over the last decades, the Minimal Supergravity (mSUGRA)~\cite{msugra} 
scenario has been extensively studied. 
In mSUGRA, symmetry breaking is achieved via gravitational 
interactions and the vast SUSY parameter space is reduced 
to only five parameters that determine the 
low energy phenomenology from the scale of Grand 
Unification(GUT): the common scalar mass $m_0$, 
the common gaugino mass $m_{1/2}$, the common soft trilinear 
SUSY breaking parameters $A_0$, the ratio of the Higgs vacuum 
expectation values at the electroweak scale tan$\beta$, 
and the sign of the Higgsino mass term sgn($\mu$).  
When R-parity~\footnote{$\rm R^{}_{P} = (-1)^{3(B-L)+2s}$, 
where B is the baryon number, L is the lepton number, and s is the spin. 
 $\rm R^{}_{P}$ is 1 for SM particles, -1 for supersymmetric particles.} 
 is conserved, supersymmetric particles have to be produced in pairs and 
ultimately decay into the lightest supersymmetric particle (LSP), usually identified 
as the lightest neutralino $\tilde{\chi}^{0}_{1}$, 
which constitutes a valid candidate for cold dark matter. 
 
At the Tevatron, the production of squarks ($\sq$) and gluinos ($\gl$), 
superpartners of quarks and gluons, constitutes one of 
the most promising channels because of the strong couplings involved. 
The cascade decay of gluinos and squarks into quarks and gluons will result 
in a final state consisting of several jets plus missing transverse 
energy (\MET) coming from the neutralinos, which leave CDF  undetected. 
In this document, we present results on a search for squarks and 
gluinos carried out on 1.1 fb$^{-1}$ of data~\footnote{At the time of writing, 
the search has been updated using 1.4 fb$^{-1}$ of data collected by the CDF 
detector.}  collected by the CDF detector in  Run II.  

\section{Signal and background samples}
\label{sec:1} 
A mSUGRA scenario with $A_0 =0$, sgn($\mu$)=-1 and tan$\beta$=5  is assumed.   
The gluino-squark mass plane is scanned via variations of the parameters 
$m_0$ (0-500 GeV/c$^{2}$) and $m_{1/2}$ (50-200 GeV/c$^{2}$).  
The {\sc pythia}~\cite{pythia} Monte Carlo program is used to generate  
samples for each mSUGRA point, and {\sc isasugra} 7.74 is implemented to 
predict the SUSY spectrum at the TeV scale. 
Light-flavor squark masses are considered degenerate, while  
2-to-2 processes involving stop ($\tilde{t}$) 
and sbottom ($\tilde{b}$) production are excluded to avoid a  
strong theoretical dependence on the mixing in the third generation.  
Each squark/gluino production subprocess (corresponding to one of the 
final states $\gl \gl$, $\sq \sq$, $\sq \bar{\sq}$, $\sq \gl$ and c.c.) 
is normalized to next-to-leading order (NLO) predictions as 
estimated using {\sc prospino v.2}~\cite{prospino}. 
The cross sections are computed using CTEQ6.1M 
PDFs and the renormalization/factorization scale is set to the average mass 
of the two final-state supersymmetric particles. 

SM backgrounds are dominated by QCD multijet processes 
where the observed $\,$ $\ETM$ $\,$ comes from partially reconstructed jets in the 
final state. $\,$ Monte Carlo samples $\,$ have been generated 
using {\sc pythia tune a}, and data in a low $\,$ $\ETM$ $\,$ region are used 
to determine the absolute normalization.  $\,$ 
Other sources of backgrounds are Z and W production in association 
with jets, and top and diboson production. For these processes, $\ETM$ might 
result from mis-reconstructed jets, the presence of neutrinos and muons in the 
final state, or both. 
{\sc alpgen} v2.1~\cite{mlm} interfaced with parton shower from {\sc pythia} is used 
to estimate W and Z+jets backgrounds, taking the normalization from the measured 
inclusive Drell-Yan cross sections~\cite{WZincl}. 
Top and diboson processes are estimated using {\sc pythia tune a} 
Monte Carlo samples normalized to the NLO theoretical predictions~\cite{top}. 
Finally, the generated samples are recontructed using the full CDF detector 
simulation.

\section{Event Selection}
\label{sec:2} 
Events are required to have a reconstructed primary vertex with 
$z$-position within 60 cm of the nominal interaction point, $\ETM>$70 GeV 
and a tracking activity consistent with the energy measured in the 
calorimeter to reject cosmics and beam-halo background.  \\
Depending on the relative masses of squarks and gluinos, 
different event topologies are expected. 
If squarks are significantly lighter than gluinos, $\sq\sq$ production 
is enhanced. The squark tends to decay according to 
$\sq \rightarrow  q \tilde{\chi}^{0}_{1}$, and a dijet topology is 
favoured, along with \MET due to the two neutralinos in the final state. 
If gluinos are lighter than squarks, $\gl\gl$ process dominates.  
Gluinos decay via $\gl \rightarrow q \bar{q} \chi^{0}_{1}$,  
leading to topologies containing a large number of jets ($\geq 4$) and 
moderate \MET. 
For $\rm M_{\tilde{g}} \approx M_{\tilde{q}}$, a topology 
with at least three jets in the final state is expected. 
Three different analyses are carried out in 
parallel, requiring at least  2, 3 or 4 jets in the final state, 
respectively. In each case, the transverse energy of the jets 
must be above 25 GeV, and one of the leading jet is 
required to be central ($|\eta | < $1.0).  
A minimum azimuthal distance between the jets and the 
\MET is required to reduce the QCD multijets contribution, and  
lepton vetoes are applied to reject boson+jets, dibosons   
and $\rm t\bar{t}$ backgrounds.   

For each final state, a dedicated study has been carried out to 
define the selection criteria that enhance the sensitivity 
to the mSUGRA signals. The signal significance, 
S/$\sqrt{\rm B}$, with S denoting the signal and B the background 
number of events, is maximized. 
Jet transverse energies, $\ETM$ and $\rm H^{}_{T}$, defined 
as the sum of the transverse energy of the jets, are the 
variables employed in the optimization.  
Table 1 summarizes the thresholds applied on 
these variables in the different analyses. 
For the 3-jet analysis, three set of thresholds, intended for
 different squark/gluino masses, are defined in order to further 
improve the signal significance across the plane where 
$\rm M_{\tilde{g}} \approx M_{\tilde{q}}$.    
\begin{table}[b]
\caption{Set of thresholds employed in the analysis.}
\label{tab:1}       
\begin{tabular}{cccccc}
\hline\noalign{\smallskip}
[GeV] & 4-jet & 3-jet(A) & 3-jet(B) &  3-jet(C) & 2-jet  \\
\noalign{\smallskip}\hline\noalign{\smallskip}
$\rm H^{}_{T}$ & 280 & 230 & 280 & 330 & 330 \\ 
\MET & 90 & 75 & 90 & 120 & 180 \\ 
$\rm E^{jet1}_{T}$ & 95 & 95 & 120 & 140 & 165 \\ 
$\rm E^{jet2}_{T}$ & 55 & 55 & 70 & 100 & 100 \\ 
$\rm E^{jet3}_{T}$ & 55 & 25 & 25 & 25 & -- \\ 
$\rm E^{jet4}_{T}$ & 25 & -- & -- & -- &  -- \\ 
\noalign{\smallskip}\hline
\end{tabular}
\vspace*{0.5cm}  
\end{table}

\section{Control regions}
\label{sec:4} 
\begin{figure}[t]
{\centerline
{\includegraphics[width=0.25\textwidth,height=0.22\textwidth,angle=0]{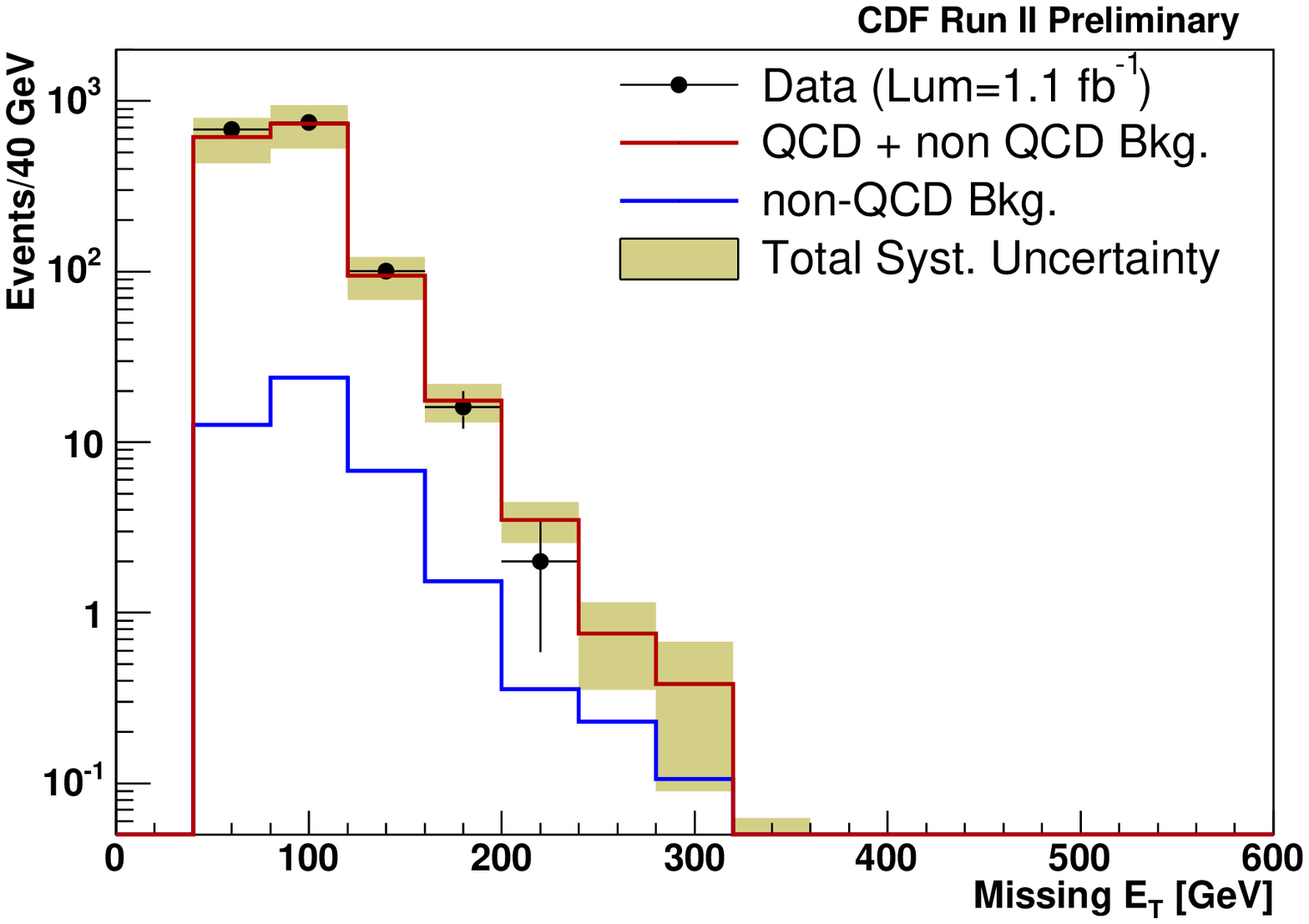}
\includegraphics[width=0.25\textwidth,height=0.22\textwidth,angle=0]{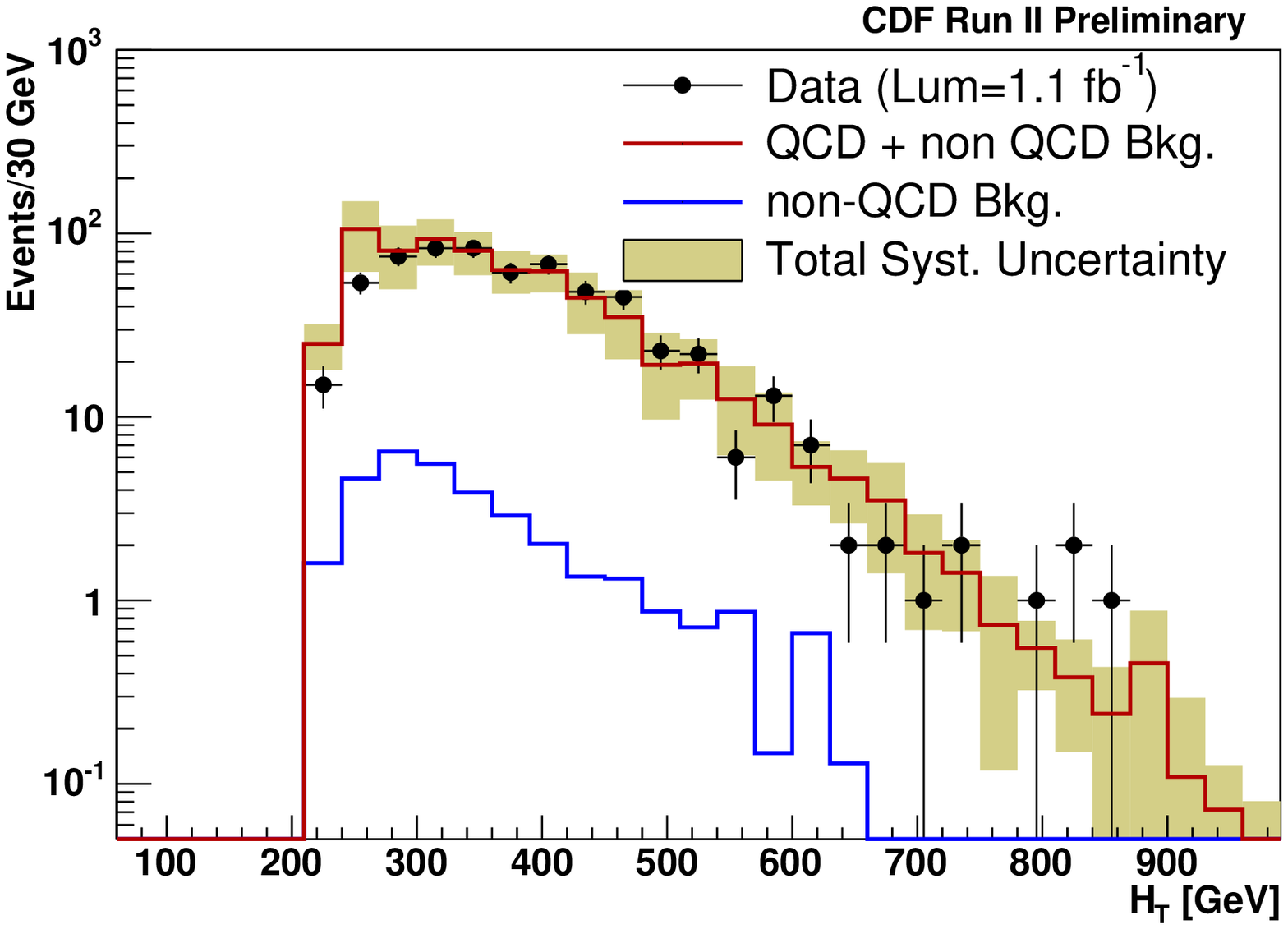}}}
\caption{$\ETM $ and $\rm H_T$ in QCD-dominated samples, after all selection cuts for the 4-jet region: 
data are superimposed to QCD$\oplus$non-QCD background with total systematics uncertainties. 
All cuts have been applied except the one on the variable that is represented. }
\label{fig:dphi_met}
\end{figure}

Monte Carlo predictions for SM processes are tested in 
background-dominated regions, referred as control regions, 
defined by reversing the selection requirements introduced 
to suppress specific background processes.     
Two different types of control regions are identified: 
\begin{itemize}
\item \emph{QCD-dominated region}: at least one of the leading 
jets is required to be aligned with the \MET azimuthal direction.  
\item \emph{EWK/Top-dominated region}: lepton vetoes are reversed, 
and at least one electron or muon, identified via the 
presence of an electromagnetic cluster or isolated track,  
is required in the event. 
\end{itemize}
Figure~\ref{fig:dphi_met} shows, for the 4-jet analysis, the 
$\ETM$ and $\rm H^{}_{T}$ distributions in the QCD-dominated sample, 
where the shaded band indicates the total systematic uncertainty on the 
background estimation (see below). Good agreement is found between 
data and SM Monte Carlo predictions.  
Similar agreement is observed in all control samples considered for 
each separately analysis.   

\section{Systematic uncertainties}
\label{sec:3} 
The dominant systematic uncertainty on the signal efficiencies 
and SM background yields comes from the 3$\%$ uncertainty on the 
jet energy scale. This translates into a 10$\%$ and 20$\%$ 
uncertainty on the signal and background estimations, respectively. 
Systematic uncertainties related to the modeling of the 
initial and final state radiation (ISR/FSR) in mSUGRA signal 
and $\rm t\bar{t}$ Monte Carlo samples translates into an 
uncertainty on the signal efficiency between 3$\%$ to 6$\%$ 
depending on the $\sq/\gl$ masses, and in a 8$\%$ uncertainty 
on the top background contribution, as defined using additional samples 
generated with modified parton showers.
Other uncertainties specific for the different background 
processes include: a 10$\%$ uncertainty on the diboson production 
cross section due to the uncertainties on PDFs and the choice of 
renormalization scale, a 2$\%$ systematic uncertainty on boson+jets production 
cross section, and a 10$\%$ uncertainty on the ${\rm t\bar{t}}$  
theoretical cross section. Finally, a 6$\%$ uncertainty on the integrated 
luminosity is also included. 

Systematic uncertainties on the mSUGRA cross sections 
are dominated by PDFs and renormalization scale uncertainties. 
The Hessian method~\cite{hessian} is used to compute 
the uncertainty due to PDFs. This translates into 
an uncertainty on the theoretical cross section that varies 
between 20$\%$ and 34$\%$, depending on the mSUGRA point.  
The uncertainty due to the renormalization and factorization scale 
is determined by re-evaluating the cross section using 
twice and half the nominal scale set in {\sc prospino} for each 
production subprocess. 
The resulting systematic 
uncertainty varies between 17$\%$ and 23$\%$, depending on the mSUGRA point.

\section{Results}
\label{sec:5} 
\begin{figure}[ht]
{\centerline
{\includegraphics[width=0.25\textwidth,height=0.22\textwidth,angle=0]{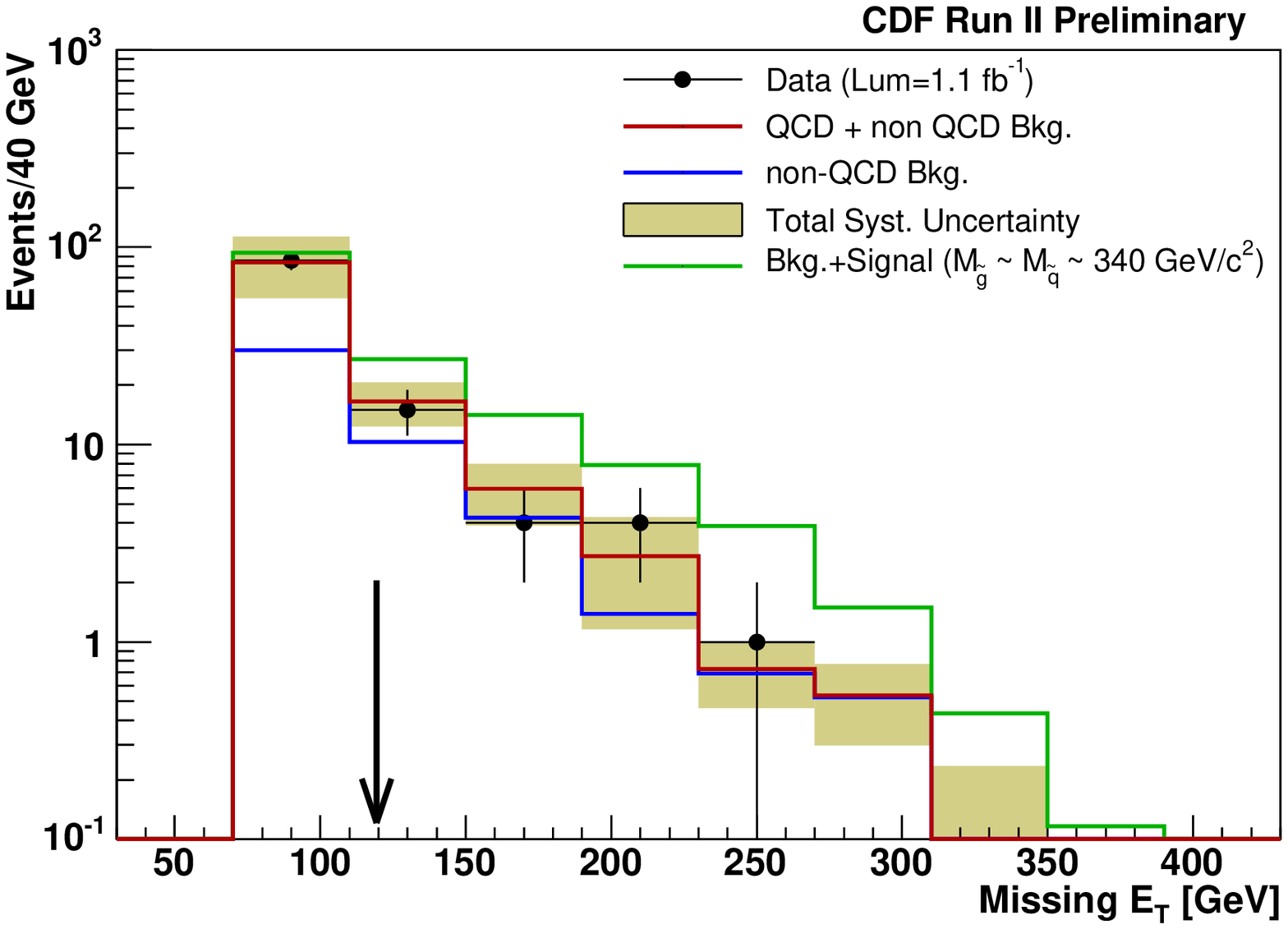}
\includegraphics[width=0.25\textwidth,height=0.22\textwidth,angle=0]{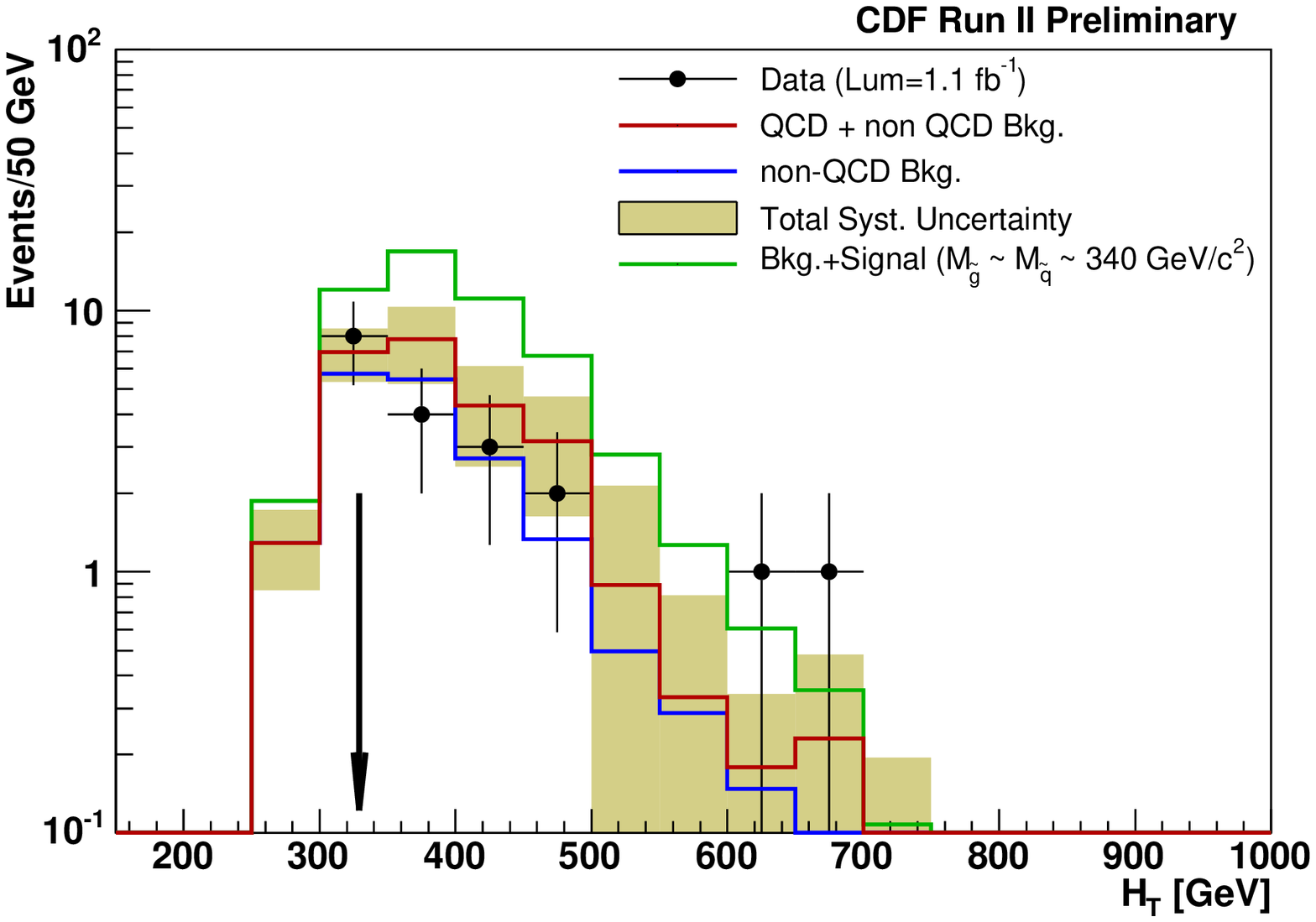}}}
{\centerline
{\includegraphics[width=0.25\textwidth,height=0.22\textwidth,angle=0]{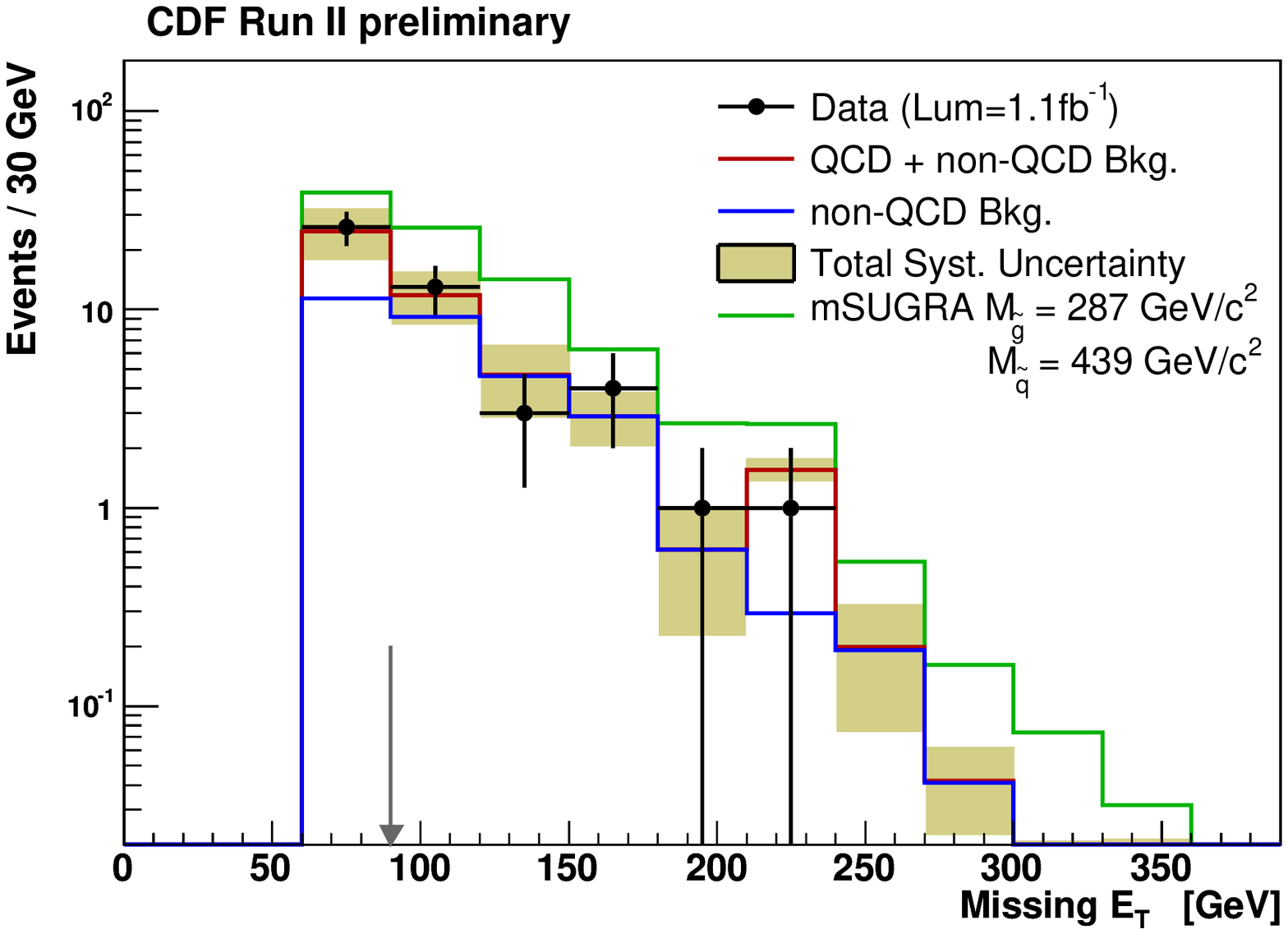}
\includegraphics[width=0.25\textwidth,height=0.22\textwidth,angle=0]{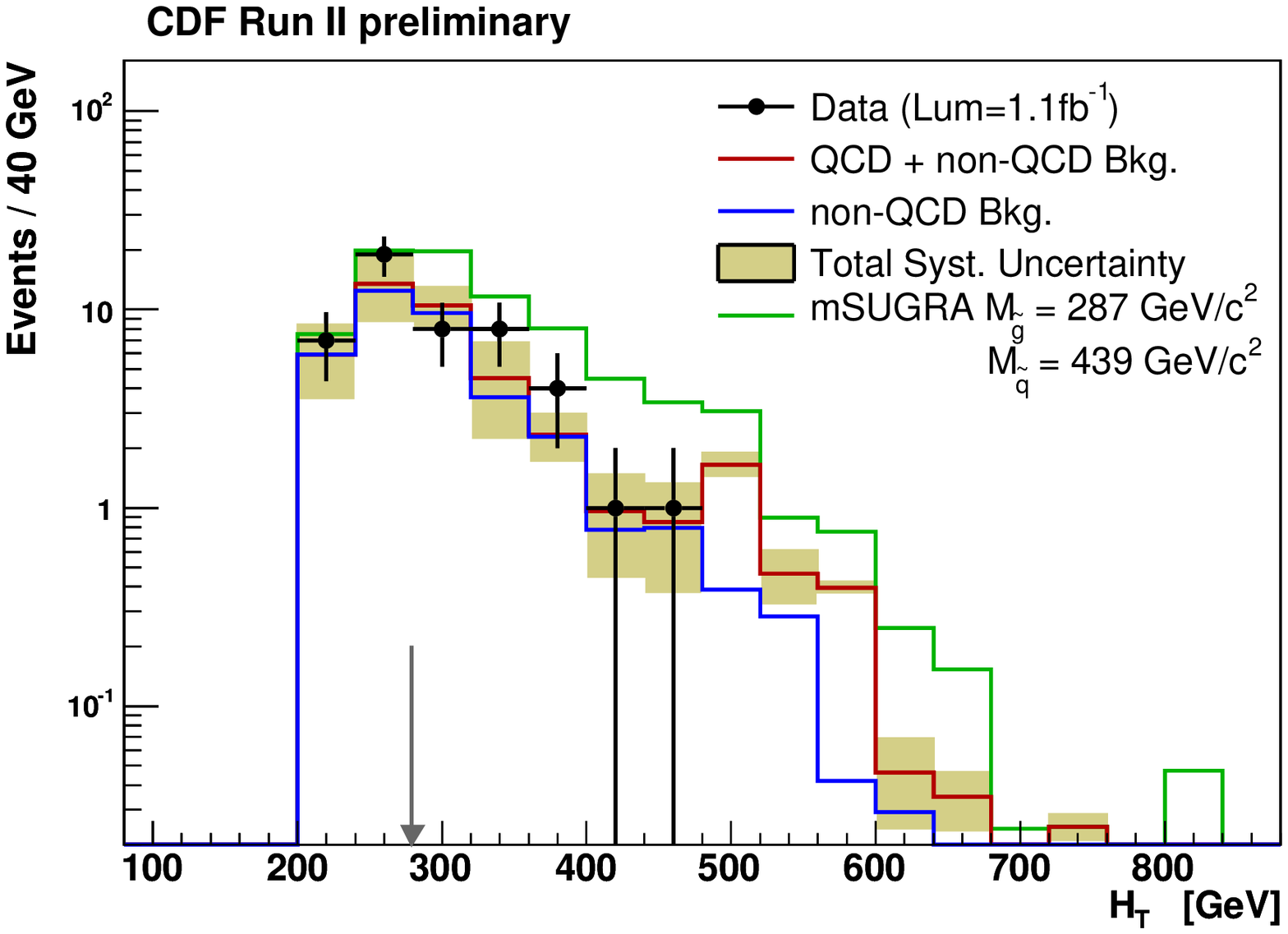}}}
\caption{$\ETM $ and $\rm H_T$ after all selection cuts for the 3-jet(C) (top) and 4-jet (bottom) analyses: 
data are superimposed to QCD and non-QCD background with total systematics uncertainties. 
Distributions for a representative signal point is shown for both 3-jet(C) and 4-jet analysis regions. 
In these plots, all cuts have been applied except the one on the variable that is represented. }
\label{fig:signal}
\end{figure}

The number of observed and SM expected events corresponding to
a total integrated luminosity of 1.1 $\rm fb^{-1}$ are reported in 
Table 2, separately for the five analyses. Good agreement is found between 
data and Monte Carlo predictions. 
Figure~\ref{fig:signal} shows the $\ETM$ and $\rm H_T$ distributions for the
3-jet(C) and 4-jet analyses. All selection criteria have been applied 
except the one on the variable that is represented. Each Figure shows the 
data superimposed on the expected SM background and the total systematic 
uncertainty. For illustration purposes, one representative mSUGRA  
point is shown for each analysis region. In case of the 3-jet(C) region, 
a point with M$_{\tilde{\rm g}} \approx \rm M_{\tilde{\rm q}} \approx 340$ GeV/c$^{2}$ 
is chosen. For the 4-jet region, a point with 
M$_{\tilde{\rm g}}=287$ GeV/c$^{2}$ and M$_{\tilde{\rm q}}=439$ GeV/c$^{2}$ 
is considered.   

\begin{figure}[ht]
\includegraphics[width=0.5\textwidth,height=0.3\textwidth,angle=0]{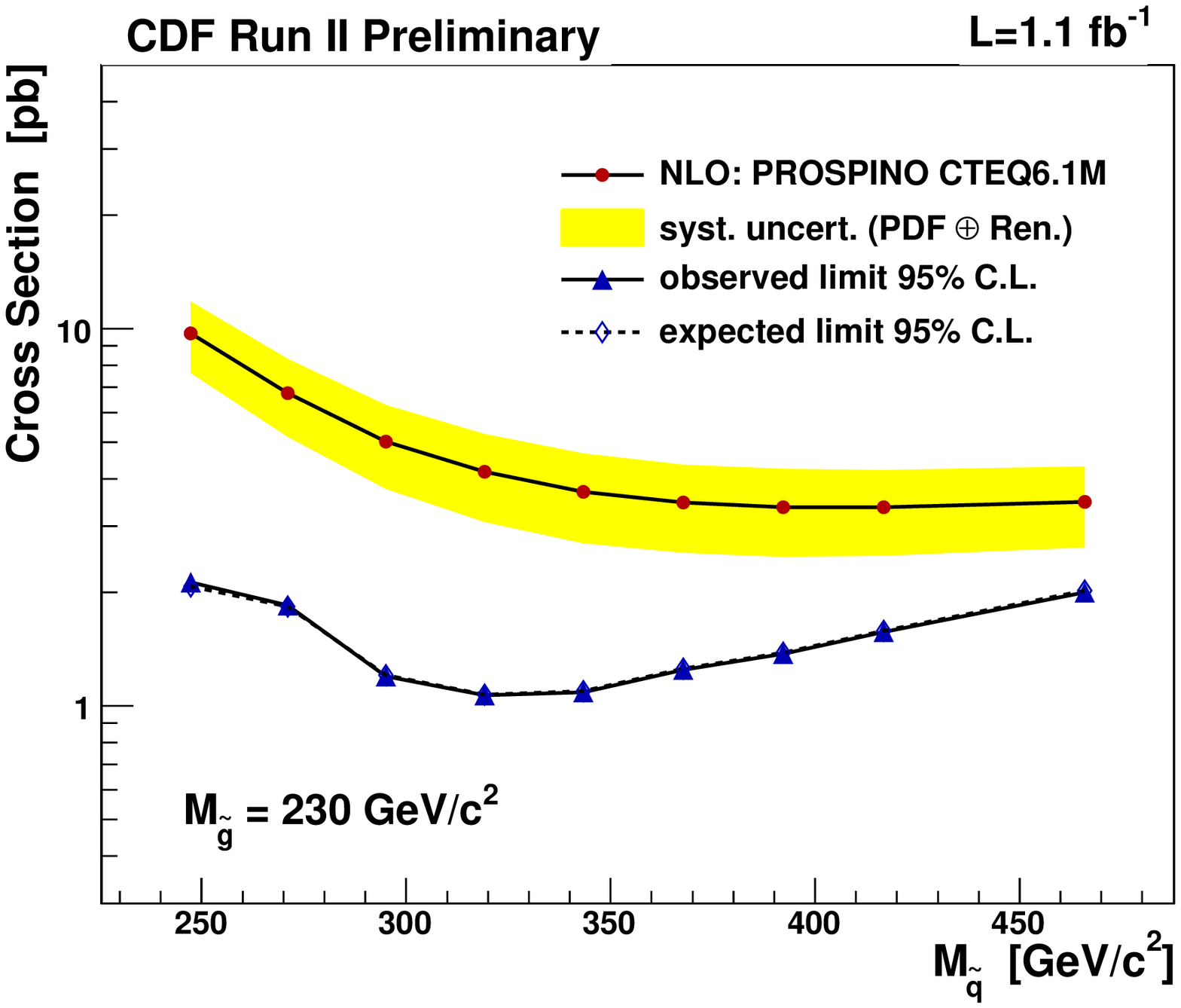}
\includegraphics[width=0.5\textwidth,height=0.3\textwidth,angle=0]{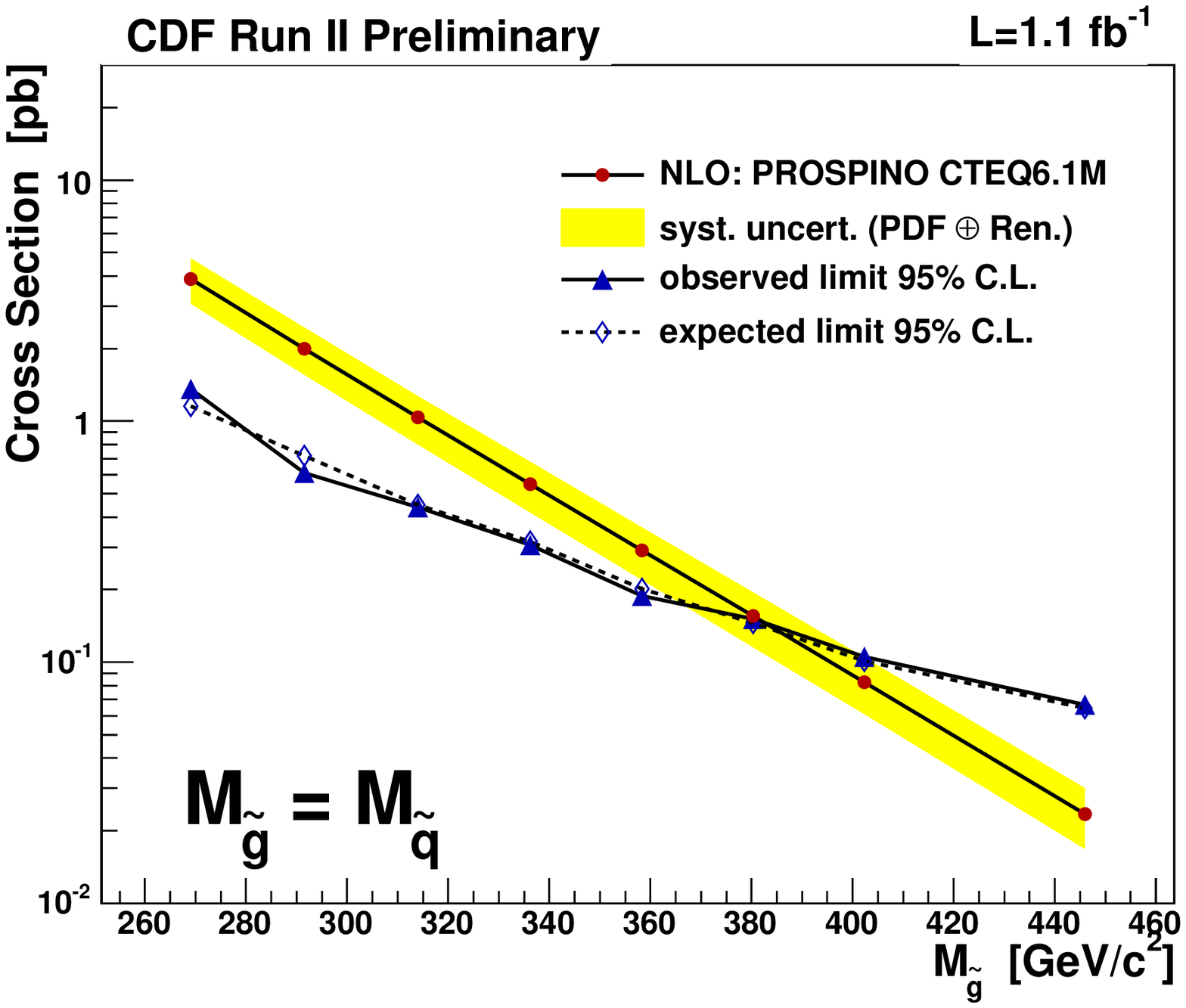}
\caption{Cross section as a function of gluino/squark masses in the case of $\rm M_{\tilde{\rm g}} = 230$ GeV/c$^{2}$ (top) 
and M$_{\tilde{\rm q}} \approx \rm M_{\tilde{\rm g}}$ (bottom). 
The observed and expected limits at $95\%$ C.L. are also shown. 
The yellow band indicates the total systematic uncertainties on the NLO prediction. 
In case of $\rm M_{\tilde{\rm g}} \approx 230$ GeV/c$^{2}$, squark masses are excluded up to 
arbitrarily high values. In case M$_{\tilde{\rm q}} \approx \rm M_{\tilde{\rm g}}$, 
squark and gluino masses are excluded up to 385 GeV/c$^{2}$.}
\label{fig:cross_1}       
\end{figure}

Since no significant deviation from the SM predictions is observed, 
results are translated into exclusion limits on gluino and squark production 
as a function of gluino and squark masses. A Bayesian approach at 
95$\%$ C.L. is used, where statistical and systematic 
uncertainties are included in the limit calculation taking into account 
correlations between signal and background uncertainties.   
The uncertainties on the NLO signal cross sections are also included.  \\
Figure~\ref{fig:cross_1} shows the {\sc prospino} NLO cross section as a funtion of the 
squark masses for M$_{\tilde{\rm g}} \approx 230$ GeV/c$^{2}$,   
and as a function of the gluino mass for M$_{\tilde{\rm g}} \approx \rm M_{\tilde{\rm q}}$. 
The yellow band indicates the total systematic uncertainties on the NLO predictions.  

\begin{figure}[ht]
\includegraphics[width=0.5\textwidth,height=0.5\textwidth,angle=0]{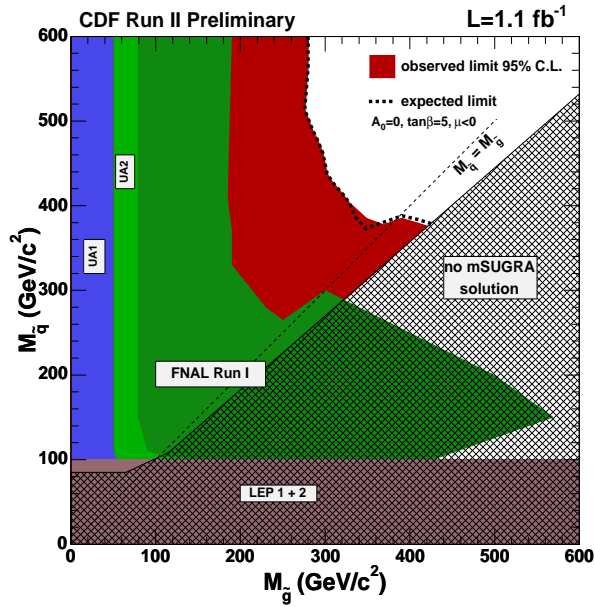}
\caption{Observed(red region) and expected (dashed-line) exclusion limits at the 95 $\%$ C.L 
in the M$_{\tilde{\rm q}}$ - M$_{\tilde{\rm g}}$ mass plane. Regions excluded by previous experiments 
are shown, together with no-mSUGRA solution region. } 
\label{fig:final_limit1}
\end{figure}

The figure also shows the curves for expected and observed limits,  
where the crossing point with the nominal NLO prediction 
would define the maximum values for squark/gluino masses 
as resulting by this analysis.    
The obtained exclusion limits from different squark and gluino 
masses are mapped out in Figure~\ref{fig:final_limit1} where  
exclusion regions, as determined by other experiments, are also presented. \\
This search excludes masses up to $385 \,  ~\mGeVcc$ at 95\% C.L. in the 
region where gluino and squark masses are similar, gluino masses up 
to $280 \, ~\mGeVcc$ for every squark mass, and gluino masses up 
to $410 \, ~\mGeVcc$ for squark masses below $380 \, ~\mGeVcc$, in a 
mSUGRA scenario with $A_0=0$, $\mu<0$ and $\tan\beta=5$.  

\begin{figure}[ht]
\includegraphics[width=0.5\textwidth,height=0.35\textwidth,angle=0]{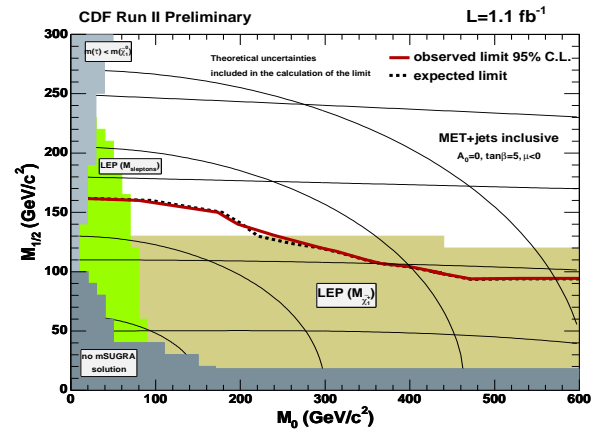}
\caption{Observed(solid line) and expected (dashed line) exclusion limits at the 95$\%$ C.L. 
in the ($m_{0}$,$m_{1/2}$) plane. In the dark-grey region there is no mSUGRA solution. 
The light-grey region indicates the region where $m(\tilde{\tau}) < m(\tilde{\chi}^{0}_{1})$. 
The beige and green regions are excluded by LEP2 chargino and slepton searches\cite{lep}, respectively. 
No mSUGRA solution is found in the dark-grey region. 
The light-grey region indicates the region 
where $m(\tilde{\tau}) < m(\tilde{\chi}^{0}_{1})$. 
The black lines are the iso-mass 
curves for gluinos (horizontal) and squarks (elliptic)  
corresponding to masses of 150, 300, 450 and 600 GeV/c$^{2}$. }
\label{fig:final_limit4}
\end{figure}

The results of this analysis in terms of M$_{\tilde{\rm g}}$-$\rm M_{\tilde{\rm q}}$ 
can also be translated into the mSUGRA parameters at the GUT scale. 
Figure~\ref{fig:final_limit4} shows the excluded regions in the 
$m^{}_{0}$-$m^{}_{1/2}$ plane for tan$\beta$=5, $A^{}_{0}$=0 and $\mu<$0. 
This search improves on the limits from indirect searches 
as determined by LEP2~\cite{lep} for $m^{}_{0}$ values between 
75 and 250 GeV/c$^{2}$ and for $m^{}_{1/2}$ values between 130 and 165 GeV/c$^{2}$.  
%
\begin{table}
\caption{Observed number of data events for the five 
selection analyses in 1.1 fb$^{-1}$, compared with the 
expected events from SM processes. 
The quoted systematic uncertainty 
on the background include 6$\%$ uncertainty on the luminosity.   
}
\label{tab:2}       
\begin{tabular}{lcc}
\hline\noalign{\smallskip}
Region & DATA & SM Expected   \\
\noalign{\smallskip}\hline\noalign{\smallskip}
4-jets    & 22 & 22 $\pm$ 2(stat.)$\pm$ 7(syst.)  \\
3-jets(A) & 494 & 484 $\pm$ 14(stat.) $\pm$ 117(syst.) \\
3-jets(B) & 136 & 123 $\pm$ 6(stat.) $\pm$ 35(syst.) \\
3-jets(C) & 17 &  20 $\pm$ 2(stat.) $\pm$ 6(syst.) \\
2-jets    & 9 &  9 $\pm$ 1(stat.) $\pm$ 2(syst.)\\
\noalign{\smallskip}\hline
\end{tabular}
\vspace*{0.5cm}  
\end{table}

\section{Conclusion}
\label{sec:concl} 
We have presented preliminary results on searches for squarks and gluinos 
in proton-antiproton collisions with a center-of-mass energy of 1.96 TeV at the Tevatron, based on 
1.1 fb$^{-1}$ of data collected by the CDF detector in  Run II.
In a mSUGRA scenario with $A_0=0$, $\mu<0$ and $\tan\beta=5$, we exclude 
masses up to about $385~\mGeVcc$ at 95\% C.L. in the region where gluino 
and squark masses are similar, gluino masses up to $280~\mGeVcc$ for 
every squark mass, and gluino masses up to $410~\mGeVcc$ for squark masses 
below $380~\mGeVcc$.   
In the future, Tevatron is expected to deliver a total integrated luminosity 
of about 6 fb$^{-1}$. This will allow either a discovery or further 
improvement on the current limits on squark and gluino masses.   

%
%
%

\end{document}